\documentclass[twocolumn,preprintnumbers,amsmath,amssymbm,prd]{revtex4}
\usepackage{epsfig}
\usepackage{graphicx}
\usepackage{amssymb}

\begin{document}
\title{Resonance spectra of caged black holes}
\author{Shahar Hod}
\affiliation{The Ruppin Academic Center, Emeq Hefer 40250, Israel}
\affiliation{ } \affiliation{The Hadassah Institute, Jerusalem
91010, Israel}
\date{\today}

\begin{abstract}
\ \ \ Recent numerical studies of the coupled Einstein-Klein-Gordon
system in a cavity have provided compelling evidence that {\it
confined} scalar fields generically collapse to form black holes.
Motivated by this intriguing discovery, we here use analytical tools
in order to study the characteristic resonance spectra of the
confined fields. These discrete resonant frequencies are expected to
dominate the late-time dynamics of the coupled black-hole-field-cage
system. We consider caged Reissner-Nordstr\"om black holes whose
confining mirrors are placed in the near-horizon region
$x_{\text{m}}\equiv (r_{\text{m}}-r_+)/r_+\ll\tau\equiv
(r_+-r_-)/r_+$ (here $r_{\text{m}}$ is the radius of the confining
mirror and $r_{\pm}$ are the radii of the black-hole horizons). We
obtain a simple analytical expression for the fundamental
quasinormal resonances of the coupled black-hole-field-cage system:
$\omega_n=-i2\pi T_{\text{BH}}\cdot n[1+O(x^n_{\text{m}}/\tau^n)]$,
where $T_{\text{BH}}$ is the temperature of the caged black hole and
$n=1,2,3,...$ is the resonance parameter.
\end{abstract}
\bigskip
\maketitle


\section{Introduction}

Caged black holes \cite{Notecag} have a long and broad history in
general relativity. These composed objects were extensively studied
in the context of black-hole thermodynamics
\cite{Haw,Dav,Hut,Gib,York,Bro,Ben,Ch}. In addition, the physics of
caged black holes was studied with relation to the black-hole bomb
mechanism of Press and Teukolsky
\cite{PressTeu2,CarDias,Hodhs,Dego,Hodch,Li}.

Recently there is a renewed interest in the physics of caged black
holes. This renewed interest stems from the important work of
Bizo\'n and Rostworowski \cite{BizRos} who revealed that
asymptotically anti-de Sitter (AdS) spacetimes are nonlinearly
unstable. In particular, it was shown in \cite{BizRos} that the
dynamics of massless, spherically-symmetric scalar fields in
asymptotically AdS spacetimes generically leads to the formation of
Schwarzschild-AdS black holes.

It is well-known that the AdS spacetime can be regarded as having an
infinite potential wall at asymptotic infinity \cite{Noteads}. One
therefore expects the dynamics of {\it confined} scalar fields
\cite{Notecos} to display a qualitatively similar behavior to the
one observed in \cite{BizRos}. In an elegant work, Maliborski
\cite{Mal} (see also \cite{Ocp,Wit}) has recently confirmed this
physically motivated expectation. In particular, the recent
numerical study by Okawa, Cardoso and Pani \cite{Ocp} provides
compelling evidence that spherically-symmetric confined scalar
fields generically collapse to form caged black holes.

The late-time dynamics of perturbation fields in a black-hole
spacetime \cite{Notelate} is characterized by quasinormal ringing,
damped oscillations which reflect the dissipation of energy from the
black-hole exterior region (see \cite{Nol,Ber,Kon} for excellent
reviews and detailed lists of references). The observation of these
characteristic complex resonances may allow one to determine the
physical parameters of the newly born black hole.

While there is a vast literature on the quasinormal spectra of black
holes in asymptotically AdS spacetimes \cite{Nol,Ber,Kon}, much less
is known about the corresponding resonances of caged black holes
\cite{Noteemp}. The recent interest \cite{Mal,Ocp,Wit} in the
dynamics and formation of caged black holes makes it highly
important to study their characteristic resonance spectra.
As we shall show below, the resonant frequencies of caged black
holes can be determined {\it analytically} in the regime
\begin{equation}\label{Eq1}
{{r_{\text{m}}-r_+}\over{r_+-r_-}}\ll1\
\end{equation}
of ``tightly caged black holes" \cite{Notetig}. Here $r_{\text{m}}$
is the radius of the confining cage (mirror) and $r_{\pm}$ are the
radii of the black-hole horizons [see Eq. (\ref{Eq4}) below].

\section{Description of the system}

The physical system we explore consists of a massless scalar field
$\Psi$ linearly coupled to a Reissner-Nordstr\"om (RN) black hole of
mass $M$ and electric charge $Q$. In terms of the Schwarzschild
coordinates $(t,r,\theta,\phi)$, the black-hole spacetime is
described by the line element \cite{Chan}
\begin{equation}\label{Eq2}
ds^2=-f(r)dt^2+{1\over{f(r)}}dr^2+r^2(d\theta^2+\sin^2\theta
d\phi^2)\ ,
\end{equation}
where \cite{Noteunit}
\begin{equation}\label{Eq3}
f(r)\equiv 1-{{2M}\over{r}}+{{Q^2}\over{r^2}}\  .
\end{equation}
The radii of the black-hole (event and inner) horizons are
determined by the zeros of $f(r)$:
\begin{equation}\label{Eq4}
r_{\pm}=M\pm (M^2-Q^2)^{1/2}\  .
\end{equation}

The dynamics of the scalar field $\Psi$ in the RN spacetime is
governed by the Klein-Gordon wave equation
\begin{equation}\label{Eq5}
\nabla^\mu\nabla_{\mu}\Psi=0\  .
\end{equation}
Resolving the field $\Psi$ into spherical harmonics:
\begin{equation}\label{Eq6}
\Psi(t,r,\theta,\phi)=\sum_{lm}Y_{lm}(\theta,\phi)R_{lm}(r)e^{-i\omega
t}/r\ ,
\end{equation}
one obtains a Schr\"odinger-like wave equation for the radial part
of the field \cite{Chan,Hsh,Notelmb}:
\begin{equation}\label{Eq7}
{{d^2R}\over{dy^2}}+\big[\omega^2-V(r)\big]R=0\  ,
\end{equation}
where the ``tortoise" radial coordinate $y$ is defined by
\begin{equation}\label{Eq8}
dy={{dr}\over{f(r)}}\  .
\end{equation}
The effective scattering potential in (\ref{Eq7}) is given by
\begin{equation}\label{Eq9}
V[r(y)]=f(r)\Big({{\lambda}\over{r^2}}+{{2M}\over{r^3}}-{{2Q^2}\over{r^4}}\Big)\
\ \ ; \ \ \ \lambda\equiv l(l+1)\  .
\end{equation}

\section{Boundary conditions}

We shall be interested in solutions of the radial wave equation
(\ref{Eq7}) with the physical requirement (boundary condition) of
purely ingoing waves crossing the black-hole horizon \cite{Chan}:
\begin{equation}\label{Eq10}
R \sim e^{-i\omega y}\ \ \ \text{as}\ \ \ r\to r_+\ \ (y\to
-\infty)\  .
\end{equation}

In addition, following \cite{Ocp} we shall consider two types of
boundary conditions at the surface $r=r_{\text{m}}$ of the confining
cavity:
\newline
(1) The Dirichlet-type boundary condition implies
\begin{equation}\label{Eq11}
R(r=r_{\text{m}})=0\  .
\end{equation}
\newline
(2) The Neumann-type boundary condition implies
\begin{equation}\label{Eq12}
{{dR}\over{dr}}(r=r_{\text{m}})=0\  .
\end{equation}

\section{The resonance conditions}

The boundary conditions (\ref{Eq11}) and (\ref{Eq12}) single out two
discrete families of complex resonant frequencies
$\{\omega(M,Q,r_{\text{m}},l;n)\}$ \cite{Noterp} which characterize
the late-time dynamics of the composed black-hole-field-cavity
system (these characteristic resonances are also known as ``boxed
quasinormal frequencies" \cite{CarDias}). The main goal of the
present paper is to determine these characteristic resonances {\it
analytically}.

Defining the dimensionless variables
\begin{equation}\label{Eq13}
x\equiv {{r-r_+}\over {r_+}}\ \ ;\ \ \tau\equiv{{r_+-r_-}\over
{r_+}}\  ,
\end{equation}
one finds [see Eqs. (\ref{Eq3}) and (\ref{Eq8})]
\begin{equation}\label{Eq14}
y={{r_+}\over{\tau}}\ln(x)+O(x)\
\end{equation}
in the near-horizon region (\ref{Eq1}), which implies \cite{Notextt}
\begin{equation}\label{Eq15}
x=e^{\tau y/r_+}[1+O(e^{\tau y/r_+})]\  .
\end{equation}
Substituting (\ref{Eq15}) into Eqs. (\ref{Eq3}) and (\ref{Eq9}) one
finds that, in the near-horizon region (\ref{Eq1}), the effective
scattering potential can be approximated by \cite{Notemq}
\begin{equation}\label{Eq16}
V(y)\to
V_{\text{near}}\equiv{{\tau(\tau+\lambda)}\over{r^2_+}}e^{\tau
y/r_+}[1+O(e^{\tau y/r_+})]\  .
\end{equation}
Substituting (\ref{Eq16}) into (\ref{Eq7}), one obtains the
Schr\"odinger-like wave equation
\begin{equation}\label{Eq17}
{{d^2R}\over{d\tilde
y^2}}+\big[\varpi^2-{{4(\tau+\lambda)}\over{\tau}}e^{2\tilde
y}\big]R=0\ ,
\end{equation}
where
\begin{equation}\label{Eq18}
\tilde y\equiv{{\tau y}\over{2r_+}}\ \ \ ; \ \ \
\varpi\equiv{{2\omega r_+}\over{\tau}}\  .
\end{equation}

Using equation 9.1.54 of \cite{Abram}, one finds that the general
solution of Eq. (\ref{Eq17}) is given by
\begin{equation}\label{Eq19}
R(z)=AJ_{-i\varpi}\big(2i\sqrt{(\tau+\lambda)/\tau}e^{\tilde y}\big)
+BJ_{i\varpi}\big(2i\sqrt{(\tau+\lambda)/\tau}e^{\tilde y}\big)\  ,
\end{equation}
where A and B are normalization constants and $J_{\nu}(x)$ is the
Bessel function of the first kind \cite{Abram}. Using equation 9.1.7
of \cite{Abram} one finds
\begin{eqnarray}\label{Eq20}
R(r\to r_+)&=&
A{{(i\sqrt{(\tau+\lambda)/\tau})^{-i\varpi}}\over{\Gamma(-i\varpi+1)}}e^{-i\omega
y}\nonumber \\
&&
+B{{(i\sqrt{(\tau+\lambda)/\tau})^{i\varpi}}\over{\Gamma(i\varpi+1)}}e^{i\omega
y}
\end{eqnarray}
for the asymptotic near-horizon ($r\to r_+$ with $e^{\tilde y}\to
0$) behavior of the radial function (\ref{Eq19}). Taking cognizance
of Eqs. (\ref{Eq10}) and (\ref{Eq20}), one concludes that the
physically acceptable solution [the one which obeys the ingoing
boundary condition (\ref{Eq10}) at the black-hole horizon] is
characterized by $B=0$. Thus, the physical solution of the radial
equation (\ref{Eq17}) is given by \cite{Noteher}
\begin{equation}\label{Eq21}
R(x)=AJ_{-i\varpi}\big(2i\sqrt{(\tau+\lambda)x/\tau}\big)\  .
\end{equation}

The Dirichlet-type boundary condition $R(x=x_{\text{m}})=0$ [see Eq.
(\ref{Eq11})] now reads
\begin{equation}\label{Eq22}
J_{-i\varpi}\big(2i\sqrt{(\tau+\lambda)x_{\text{m}}/\tau}\big)=0\ .
\end{equation}
Using equation 9.1.2 of \cite{Abram}, one can express this boundary
condition in the form
\begin{equation}\label{Eq23}
\tan(i\varpi\pi)={{J_{i\varpi}\big(2i\sqrt{(\tau+\lambda)x_{\text{m}}/\tau}\big)}\over
{Y_{i\varpi}\big(2i\sqrt{(\tau+\lambda)x_{\text{m}}/\tau}\big)}}\ ,
\end{equation}
where $Y_{\nu}(x)$ is the Bessel function of the second kind
\cite{Abram}. In the near-horizon region [see Eq. (\ref{Eq1})]
\begin{equation}\label{Eq24}
z_{\text{m}}\equiv (\tau+\lambda){{x_{\text{m}}}\over{\tau}}\ll1
\end{equation}
one may use equations 9.1.7 and 9.1.9 of \cite{Abram} in order to
write the resonance condition (\ref{Eq23}) in the form
\cite{Notecorr}
\begin{equation}\label{Eq25}
\tan(i\varpi\pi)=i{{\pi
e^{-\pi\varpi}z^{i\varpi}_{\text{m}}}\over{\varpi\Gamma^2(i\varpi)}}[1+O(z_{\text{m}})]\
.
\end{equation}

The Neumann-type boundary condition $dR(x=x_{\text{m}})/dx=0$ [see
Eq. (\ref{Eq12})] now reads
\begin{equation}\label{Eq26}
{{d}\over{dx}}\Big[J_{-i\varpi}\big(2i\sqrt{(\tau+\lambda)x/\tau}\big)\Big]_{x=x_{\text{m}}}=0\
.
\end{equation}
Using equation 9.1.27 of \cite{Abram}, one can express (\ref{Eq26})
in the form
\begin{eqnarray}\label{Eq27}
J_{-i\varpi-1}\big(2i\sqrt{(\tau+\lambda)x_{\text{m}}/\tau}\big)-
\nonumber \\
J_{-i\varpi+1}\big(2i\sqrt{(\tau+\lambda)x_{\text{m}}/\tau}\big)=0\
. \nonumber \\ &&
\end{eqnarray}
Using equation 9.1.2 of \cite{Abram}, one can express this boundary
condition in the form
\begin{equation}\label{Eq28}
\tan(i\varpi\pi)={{J_{i\varpi+1}(2i\sqrt{z_{\text{m}}})-J_{i\varpi-1}(2i\sqrt{z_{\text{m}}})}
\over{Y_{i\varpi+1}(2i\sqrt{z_{\text{m}}})-Y_{i\varpi-1}(2i\sqrt{z_{\text{m}}})}}\
.
\end{equation}
From equations 9.1.7 and 9.1.9 of \cite{Abram} one finds
$J_{i\varpi+1}(2i\sqrt{z_{\text{m}}})/J_{i\varpi-1}(2i\sqrt{z_{\text{m}}})=O(z_{\text{m}})\ll1$
and
$Y_{i\varpi+1}(2i\sqrt{z_{\text{m}}})/Y_{i\varpi-1}(2i\sqrt{z_{\text{m}}})=O(z^{-1}_{\text{m}})\gg1$
in the near-horizon $z_{\text{m}}\ll1$ region [see Eq.
(\ref{Eq24})]. Using these relations, one may write the resonance
condition (\ref{Eq28}) in the form \cite{Notecorr}
$\tan(i\varpi\pi)=-J_{i\varpi-1}(2i\sqrt{z_{\text{m}}})/Y_{i\varpi+1}(2i\sqrt{z_{\text{m}}})[1+O(z_{\text{m}})]$,
which in the near-horizon region (\ref{Eq24}) implies (see equations
9.1.7 and 9.1.9 of \cite{Abram})
\begin{equation}\label{Eq29}
\tan(i\varpi\pi)=-i{{\pi
e^{-\pi\varpi}z^{i\varpi}_{\text{m}}}\over{\varpi\Gamma^2(i\varpi)}}[1+O(z_{\text{m}})]\
.
\end{equation}

\section{The discrete resonance spectra of caged black holes}

Taking cognizance of the near-horizon condition (\ref{Eq24}), one
realizes that the r.h.s of the resonance conditions (\ref{Eq25}) and
(\ref{Eq29}) are small quantities. This observation follows from the
fact that, for damped modes with $\Im\varpi<0$ [$\Re(i\varpi)>0$]
one has $z^{i\varpi}_{\text{m}}\ll1$ in the regime (\ref{Eq24}). We
can therefore use an iteration scheme in order to solve the
resonance conditions (\ref{Eq25}) and (\ref{Eq29}).

The zeroth-order resonance equation is given by
$\tan(i\varpi^{(0)}\pi)=0$ for both the Dirichlet-type boundary
condition (\ref{Eq11}) and the Neumann-type boundary condition
(\ref{Eq12}) [see Eqs. (\ref{Eq24}), (\ref{Eq25}) and (\ref{Eq29})].
This yields the simple zeroth-order resonances
\begin{equation}\label{Eq30}
\varpi^{(0)}_n=-in\ \ \ ; \ \ \ n=1,2,3,...
\end{equation}
of the caged black holes.

Substituting (\ref{Eq30}) into the r.h.s of (\ref{Eq25}) and
(\ref{Eq29}), one obtains the first-order resonance condition
\begin{equation}\label{Eq31}
\tan(i\varpi^{(1)}_n\pi)={\mp}{{\pi(-z_{\text{m}})^{n}}\over{n\Gamma^2(n)}}\
,
\end{equation}
where the upper sign corresponds to the Dirichlet-type boundary
condition (\ref{Eq11}) and the lower sign corresponds to the
Neumann-type boundary condition (\ref{Eq12}). From (\ref{Eq31}) one
finds \cite{Notetan}
\begin{equation}\label{Eq32}
\varpi_n=-in\Big[1{\mp}{{(-z_{\text{m}})^n} \over{(n!)^2}}\Big]\ \ ;
\ \ n=1,2,3,...
\end{equation}
for the characteristic resonance spectra of caged black holes in the
regime (\ref{Eq24}).


\section{Summary and discussion}

Recent numerical studies of the Einstein-Klein-Gordon system in a
cavity \cite{Ocp,Wit} have provided compelling evidence that
confined scalar fields \cite{Notecos} generically collapse to form
(caged) black holes. Motivated by these intriguing studies, we have
explored here the late-time dynamics \cite{Notelate} of these
confined fields in the background of caged black holes.

In particular, we have studied the characteristic resonance spectra
of confined scalar fields in caged Reissner-Nordstr\"om black-hole
spacetimes. It was shown that these resonances can be derived {\it
analytically} for caged black holes whose confining mirrors are
placed in the vicinity of the black-hole horizon [that is, in the
regime $x_{\text{m}}\ll\tau$, see Eq. (\ref{Eq24})]. Remarkably, the
resonant frequencies of these caged black holes can be expressed in
terms of the Bekenstein-Hawking {\it temperature} $T_{\text{BH}}$ of
the black-hole \cite{Notetm}:
\begin{equation}\label{Eq33}
\omega_n=-i2\pi T_{\text{BH}}\cdot
n\Big\{1{\mp}{{[-(\tau+\lambda)x_{\text{m}}/\tau]^n}
\over{(n!)^2}}\Big\}\ .
\end{equation}

Note that, for spherical field configurations (the ones studied in
\cite{Mal,Ocp}), the characteristic resonant frequencies are given
by the remarkably simple linear relation \cite{Notelin}
\begin{equation}\label{Eq34}
\omega_n=-i2\pi T_{\text{BH}}\cdot
n\Big[1{\mp}{{(-x_{\text{m}})^{n}}\over{(n!)^2}}\Big]\ .
\end{equation}

Finally, it is worth mentioning other black-hole spacetimes which
share this remarkable property (that is, black-hole spacetimes which
are characterized by resonant frequencies whose imaginary parts
\cite{Noterel} scale linearly with the black-hole temperature):
\newline
(1) Near-extremal asymptotically flat Kerr black holes \cite{Hode}.
\newline
(2) Near-extremal asymptotically flat Reissner-Nordstr\"om black
holes coupled to charged scalar fields \cite{Hodpla}.
\newline
(3) Asymptotically flat charged black holes coupled to charged
scalar fields in the highly-charged regime $qQ\gg1$
\cite{Notech,Hodwkb}.
\newline
(4) The ``subtracted" black-hole geometries studied in \cite{GibCv}.
\newline
(5) Asymptotically AdS black holes in the regime $r_+\gg R$
\cite{NoteR,Horo}. This last example, together with our result
(\ref{Eq34}) for the resonant frequencies of caged black holes,
provide an elegant demonstration of the analogy, already discussed
in the Introduction, between asymptotically AdS black-hole
spacetimes and caged black-hole spacetimes.

The confining cavity (mirror) of our analysis obviously restricts
the dynamics of the fields to the near-horizon region $x\leq
x_{\text{m}}\ll\tau$. It is therefore not surprising that the
characteristic resonances (\ref{Eq33}) of these caged black holes
are determined by the surface gravity \cite{Notest} at the
black-hole horizon. The fact that the black-hole spacetimes
mentioned above \cite{Hode,Hodpla,Hodwkb,GibCv,Horo} share this same
property (namely, they are characterized by a linear scaling of
their resonances with the black-hole temperature) suggests that the
dynamics of perturbation fields in these black-hole spacetimes are
mainly determined by the near-horizon properties of these
geometries.

\bigskip
\noindent
{\bf ACKNOWLEDGMENTS}
\bigskip

This research is supported by the Carmel Science Foundation. I thank
Yael Oren, Arbel M. Ongo and Ayelet B. Lata for helpful discussions.


\bigskip

\begin{thebibliography}{99}

\bibitem{Notecag} We use the term ``Caged black holes" to describe
black holes which are confined within finite-volume cavities.

\bibitem{Haw} S. W. Hawking, Phys. Rev. D {\bf 13}, 191 (1976).

\bibitem{Dav} P. C. W. Davies, Proc. R. Soc. London {\bf A353}, 499 (1977).

\bibitem{Hut} P. Hut, Mon. Not. R. astr. Soc. {\bf 180}, 379 (1977).

\bibitem{Gib} G. W. Gibbons and M. J. Perrry, Proc. R. Soc. London {\bf A358}, 467 (1978).

\bibitem{York} J. W. York, Phys. Rev. D {\bf 33}, 2092 (1986).

\bibitem{Bro} J. Brown, E. A. Martinez, and J. W. York, Phys. Rev. Lett. {\bf 66}, 2281 (1991).

\bibitem{Ben} B. Schumacher, W. A. Miller, and W. H. Zurek, Phys.
Rev. D {\bf 46}, 1416 (1992).

\bibitem{Ch} P. S. Cust\'odioa and J. E. Horvathb, Am. J. Phys. {\bf
71}, 1237 (2003).

\bibitem{PressTeu2} W. H. Press and S. A. Teukolsky, Nature {\bf 238}, 211 (1972).

\bibitem{CarDias} V. Cardoso, O. J. C. Dias, J. P. S. Lemos and S.
Yoshida, Phys. Rev. D {\bf 70}, 044039 (2004); Erratum-ibid. D {\bf
70}, 049903 (2004).

\bibitem{Hodhs} S. Hod, Phys. Rev. D {\bf 88}, 124007 (2013)
[arXiv:1405.1045]; S. Hod, Phys. Lett. B {\bf 736}, 398 (2014).

\bibitem{Dego} J. C. Degollado, C. A. R. Herdeiro, and H. F. R\'unarsson, Phys. Rev. D {\bf 88}, 063003
(2013); J. C. Degollado and C. A. R. Herdeiro, Phys. Rev. D {\bf
89}, 063005 (2014).

\bibitem{Hodch} S. Hod, Phys. Rev. D {\bf 88}, 064055 (2013)
[arXiv:1310.6101].

\bibitem{Li} R. Li, arXiv:1404.6309.

\bibitem{BizRos} P. Bizo\'n and A. Rostworowski, Phys. Rev. Lett. {\bf 107}, 031102
(2011).

\bibitem{Noteads} It is worth mentioning that the black-hole bomb mechanism was also studied in
the context of asymptotically AdS black holes, see: V. Cardoso and
O. J. C. Dias, Phys. Rev. D {\bf 70}, 084011 (2004); O. J. C. Dias,
G. T. Horowitz and J. E. Santos, JHEP {\bf 1107} (2011) 115; O. J.
C. Dias, P. Figueras, S. Minwalla, P. Mitra, R. Monteiro and J. E.
Santos, JHEP {\bf 1208} (2012) 117; O. J. C. Dias and J. E. Santos,
JHEP {\bf 1310} (2013) 156; V. Cardoso, O. J. C. Dias, G. S.
Hartnett, L. Lehner and J. E. Santos, JHEP 1404 (2014) {\bf 183}
[arXiv:1312.5323].

\bibitem{Notecos} That is, scalar fields which are confined within finite-volume cavities.

\bibitem{Mal} M. Maliborski, Phys. Rev. Lett. {\bf 109}, 221101
(2012).

\bibitem{Ocp} H. Okawa, V. Cardoso, and P. Pani, arXiv:1409.0533.

\bibitem{Wit} H. Witek1, V. Cardoso, L. Gualtieri, C. Herdeiro, A. Nerozzi,
U. Sperhake, and M.  Zilh\~ao, J. Phys.: Conf. Ser. {\bf 229},
012072 (2010).

\bibitem{Notelate} That is, the dynamics of the fields well after the formation of the
black-hole horizon.

\bibitem{Nol} H. P. Nollert, Class. Quantum Grav. {\bf 16}, R159 (1999).

\bibitem{Ber} E. Berti, V. Cardoso and A. O. Starinets, Class. Quant. Grav. {\bf 26},
163001 (2009).

\bibitem{Kon} R. A. Konoplya and A. Zhidenko, Rev. Mod. Phys. {\bf 83}, 793 (2011).

\bibitem{Noteemp} It is worth emphasizing again that caged black
holes may serve as a simple toy-model for the physically more
realistic AdS black holes.

\bibitem{Notetig} We use the term ``Tightly caged black holes" to reflect
the fact that the boundary of the confining cavity is placed in the
vicinity of the black-hole horizon: $r_{\text{m}}-r_+\ll r_+-r_-$.

\bibitem{Chan} S. Chandrasekhar, {\it The Mathematical Theory of Black
Holes}, (Oxford University Press, New York, 1983).

\bibitem{Noteunit} We use natural units in which $G=c=\hbar=1$.

\bibitem{Hsh} S. Hod and T. Piran, Phys. Rev. D {\bf 58},
024017 (1998) [arXiv:gr-qc/9712041]; S. Hod and T. Piran, Phys. Rev.
D {\bf 58}, 024018 (1998) [arXiv:gr-qc/9801001]; S. Hod and T.
Piran, Phys. Rev. D {\bf 58}, 024019 (1998) [arXiv:gr-qc/9801060];
T. Hartman, W. Song, and A. Strominger, JHEP 1003:118 (2010); S.
Hod, Class. Quant. Grav. {\bf 23}, L23 (2006) [arXiv:gr-qc/0511047].

\bibitem{Notelmb} We shall henceforth omit the indices $l$ and $m$ for brevity.

\bibitem{Noterp} The integer $n$ is the resonance parameter.

\bibitem{Notextt} Note that the near-horizon region (\ref{Eq1}) corresponds to
$x\leq x_{\text{m}}\ll\tau\leq1$. This also implies [see Eq.
(\ref{Eq14})] $y\to -\infty$ (and thus $e^{\tau y/r_+}\to 0$) in the
region (\ref{Eq1}).

\bibitem{Notemq} Here we have used the relation
$(\lambda+2M/r-2Q^2/r^2)/r^2=(\lambda+\tau)/r^2_+[1+O(x)]$ in the
near-horizon region (\ref{Eq1}).

\bibitem{Abram} M. Abramowitz and I. A. Stegun, {\it Handbook of
Mathematical Functions} (Dover Publications, New York, 1970).

\bibitem{Noteher} Here we have used the relation (\ref{Eq15}) in the near-horizon region (\ref{Eq1}).

\bibitem{Notecorr} See equations 9.1.10 and 9.1.11 of \cite{Abram} for
the sub-leading correction terms.

\bibitem{Notetan} Here we have used the relation $\tan(x+n\pi)=\tan(x)$. In addition,
we have used the relation $\tan(x)=x+O(x^3)$ in the $x\ll1$ regime
[see Eq. (\ref{Eq24})].

\bibitem{Notetm} Here we have used Eqs. (\ref{Eq18}) and
(\ref{Eq32}) together with the relation $T_{\text{BH}}=\tau/4\pi
r_+$ for the Bekenstein-Hawking temperature of the black hole.

\bibitem{Notelin} That is, the resonant frequencies scale {\it
linearly} with the black-hole temperature $T_{\text{BH}}$.

\bibitem{Noterel} It is worth emphasizing that, the characteristic relaxation time
of generic field perturbations is determined by the (reciprocal of
the) imaginary part of the fundamental $(n=1)$ resonance:
$\tau_{\relax}=1/\Im{\omega_1}$.

\bibitem{Hode} S. Hod, Phys. Rev. D 75, 064013 (2007) [arXiv:gr-qc/0611004]; S.
Hod, Class. and Quant. Grav. 24, 4235 (2007) [arXiv:0705.2306]; A.
Gruzinov, arXiv:gr-qc/0705.1725; S. Hod, Phys. Lett. B 666 483
(2008) [arXiv:0810.5419]; S. Hod, Phys. Rev. D 78, 084035 (2008)
[arXiv:0811.3806]; S. Hod, Phys. Rev. D 80, 064004 (2009)
[arXiv:0909.0314].

\bibitem{Hodpla} S. Hod, Phys. Lett. A {\bf 374}, 2901 (2010)
[arXiv:1006.4439].

\bibitem{Notech} Here $q$ is the charge coupling constant of the field.

\bibitem{Hodwkb} S. Hod, Phys. Lett. B {\bf 710}, 349 (2012) [arXiv:1205.5087];
R. A. Konoplya and A. Zhidenko, Phys. Rev. D {\bf 88}, 024054
(2013).

\bibitem{GibCv} M. Cvetic and G. W. Gibbons, Phys. Rev. D {\bf 89}, 064057 (2014).

\bibitem{NoteR} Here $R$ is the AdS radius.

\bibitem{Horo} G. T. Horowitz and V. E. Hubeny, Phys. Rev. D {\bf 62}, 024027
(2000).

\bibitem{Notest} Note that the surface gravity is proportional to the
black-hole temperature.

\end{thebibliography}
\end{document}